\documentclass[aps,prd,nofootinbib,amsmath,amssymb,superscriptaddress,onecolumn,11pt]{revtex4}%superscriptaddress,showpacs,

\pdfoutput=1

\usepackage{txfonts}
\usepackage{graphicx}
\usepackage{dcolumn}
\usepackage{bm}
\usepackage{amssymb}
\usepackage{latexsym}
\usepackage{booktabs}
\usepackage{amsmath}
\usepackage{multirow}
\usepackage[colorlinks=true, linkcolor=red, citecolor=blue]{hyperref}

\newcommand{\be}{\begin{equation}}
\newcommand{\ee}{\end{equation}}
\newcommand{\bq}{\begin{eqnarray}}
\newcommand{\eq}{\end{eqnarray}}

\begin{document}

\title{Constraining dark energy with Hubble parameter measurements: an analysis including future redshift-drift observations}

\author{Rui-Yun Guo}
\affiliation{Department of Physics, College of Sciences, Northeastern University, Shenyang
110004, China}
\author{Xin Zhang\footnote{Corresponding author}}
\email{zhangxin@mail.neu.edu.cn} \affiliation{Department of Physics, College of Sciences,
Northeastern University, Shenyang 110004, China}
\affiliation{Center for High Energy Physics, Peking University, Beijing 100080, China}

\begin{abstract}

The nature of dark energy affects the Hubble expansion rate (namely, the expansion history) $H(z)$ by an integral over $w(z)$. However, the usual observables are the luminosity distances or the angular diameter distances, which measure the distance-redshift relation. Actually, the property of dark energy affects the distances (and the growth factor) by a further integration over functions of $H(z)$. Thus the direct measurements of the Hubble parameter $H(z)$ at different redshifts are of great importance for constraining the properties of dark energy. In this paper, we show how the typical dark energy models, for example, the $\Lambda$CDM, $w$CDM, CPL, and holographic dark energy models, can be constrained by the current direct measurements of $H(z)$ (31 data used in total in this paper, covering the redshift range of $z\in [0.07,2.34]$). In fact, the future redshift-drift observations (also referred to as the Sandage-Loeb test) can also directly measure $H(z)$ at higher redshifts, covering the range of $z\in [2,5]$. We thus discuss what role the redshift-drift observations can play in constraining dark energy with the Hubble parameter measurements. We show that the constraints on dark energy can be improved greatly with the $H(z)$ data from only a 10-year observation of redshift drift.

\end{abstract}
%\pacs{95.36.+x, 98.80.Es, 98.80.-k}
\maketitle

\section{Introduction}
\label{sec:intro}

In 1998, two observation teams independently found that the universe is currently undergoing an accelerating expansion, through the observations of type Ia supernovae~\cite{SN,SN0}. Though the statistical significance was not high enough, the supernovae evidence for cosmic acceleration was quickly accepted by the community at large because the subsequent observations of cosmic microwave background (CMB)~\cite{Spergel:2003cb,Bennett:2003bz} and large-scale structure (LSS)~\cite{Tegmark:2003ud,Abazajian:2004aja} soon provided substantial independent evidence supporting the conclusion of supernovae observations. If the theory of general relativity (GR) is valid on all scales of the universe, the fact of cosmic acceleration implies that a new energy component with negative pressure, referred to as ``dark energy''~\cite{Sahni:1999gb,Padmanabhan:2002ji,Peebles:2002gy,Copeland:2006wr,Sahni:2006pa,Frieman:2008sn,Li:2011sd,Bamba:2012cp,Weinberg:2012es,Mortonson:2013zfa}, is needed in the universe. However, there still exists another possibility: that the cosmic acceleration arises from a breakdown of GR on cosmological scales. To distinguish between dark energy and modified gravity (MG) is a major mission in modern cosmology. A basic strategy is to accurately measure the both histories of cosmic expansion and growth of structure and to compare them for a consistency check.

The main property of dark energy is characterized by its equation-of-state parameter (EoS) $w(z)$. In fact, dark energy affects the expansion history and growth of structure of the universe in a subtle way. To measure the history of the cosmic expansion, the most important way is to measure the distance-redshift relation. For example, through the observations of type Ia supernovae, one measures the luminosity distances at different redshifts, and through the observations of baryon acoustic oscillations (BAO), one measures the angular diameter distances at different redshifts. The cosmic distance, whether the luminosity distance or the angular diameter distance, is linked to the Hubble expansion rate $H(z)$ through an integration, namely, $D_c(z)=\int_0^z dz'/H(z')$, where $D_c(z)$ is the comoving line-of-sight distance to an object at redshift $z$ in a flat universe. The luminosity distance $D_L(z)$ and the angular diameter distance $D_A(z)$ can be expressed as $D_L=(1+z)D_c$ and $D_A=(1+z)^{-1}D_c$, respectively. In fact, the linear growth factor also involves a further integration over a function of $H(z)$.

Furthermore, the property of dark energy affects the Hubble expansion rate $H(z)$ also through an integral, namely, in a flat universe, we have
\begin{equation}\label{1}
    \frac{H^{2}(z)}{H^{2}_{0}}=\Omega_{r}(1+z)^{4}+\Omega_{m}(1+z)^{3}+(1-\Omega_{r}-\Omega_{m})X(z),
\end{equation}
where $\Omega_{r}$ and $\Omega_{m}$ are the current density parameters of radiation and matter, respectively, and $X(z)$ describes how dark energy density evolves with redshift,
\begin{equation}
X(z) \equiv \rho_{\rm de}(z) / \rho_{\rm de}(0) = \exp \left[3 \int^{z}_{0} \frac{1+w(z^{\prime})}{1+z^{\prime}} dz^{\prime}\right].
\end{equation}
Therefore, it is extremely difficult to constrain the property of dark energy using the measurements of cosmic distances and growth rate of structure, because there are two integrals between these observables and $w(z)$. Obviously, to accurately constrain the history of dark energy evolution, a more important way is to directly measure the Hubble parameter $H(z)$, owing to the fact that between $H(z)$ and $w(z)$ there is only one integral. While difficult, a number of measurement data of $H(z)$ have been accumulated and studied in recent years~\cite{H1,H2,H3,Daly:2005iu,H4,MiraldaEscude:2009uz,H5,Kazin:2010nd,H7,H6,Cabre:2010bc,Chen:2011ys,H8,H9,Suzuki:2012kj,H10,Busca:2012bu,H12,Farooq:2012ju,Farooq:2013hq,Farooq:2013eea,H13,Anderson:2013oza,Font-Ribera:2013wce,Anderson:2013zyy,Li:2014yza,Sahni:2014ooa,H15,Delubac:2014aqe,Moresco:2015cya,Ding:2015vpa,H16}.

Through two astrophysical methods, namely, the measurement of differential age of galaxies and the measurement of clustering of galaxies or quasars, more than 30 observational data of $H(z)$ have been obtained~\cite{H2,H4,H5,H8,H9,Busca:2012bu,H10,Anderson:2013oza,Font-Ribera:2013wce,Anderson:2013zyy,H13,Delubac:2014aqe,Moresco:2015cya}. One of the major aims of this paper is to have a look at how these $H(z)$ data can constrain dark energy. We perform such an analysis by taking several typical dark energy models as examples. We only focus on the expansion history of the universe, thus we do not consider MG models in this paper. Since the current observations show that the spatial curvature of the universe is very small, $|\Omega_k|\lesssim {\cal O}(10^{-3})$~\cite{Ade:2015xua}, we only consider a flat universe in the analysis of this paper.

The current data of $H(z)$ are all in the range of $z\lesssim 2$. Obviously, measuring $H(z)$ at higher redshifts could provide additional accurate information as regards $\Omega_m h^2$, thus helping break the low-redshift parameter degeneracies, which is of great importance to constrain the property of dark energy. Recently, there have been a number of works discussing the observations of redshift drift~\cite{Corasaniti:2007bg,Balbi:2007fx,Zhang:2007zga,Liske:2008ph,Zhang:2010im,Quercellini:2010zr,Martinelli:2012vq,Li:2013oba,Zhang:2013zyn,sl4,sl1,sl5,sl3,sl2}, which probe the expansion history of the universe in the ``redshift desert'' of $2\lesssim z\lesssim 5$. Through monitoring the shift of Lyman-$\alpha$ forest absorption line of a distant quasar over a period of a few decades, one can detect the time variation of its redshift, namely, the redshift drift. This is equivalent to measure the Hubble parameter at a high redshift. This method is also referred to as the ``Sandage-Loeb test'' (SL test)~\cite{sandage,Loeb:1998bu}. The highly accurate COsmic Dynamics EXperiment (CODEX) spectrograph on the $39$m Extremely Large Telescope (ELT) being built is expected to perform such a task~\cite{Liske:2008ph}. The forecast analyses of using the redshift-drift observations to constrain dark energy have been recently done in a number of work~\cite{Corasaniti:2007bg,Zhang:2007zga,Balbi:2007fx,Liske:2008ph,Zhang:2010im,Quercellini:2010zr,Martinelli:2012vq,Li:2013oba,Zhang:2013zyn,sl4,sl1,sl5,sl3,sl2}. The combination of SL test data and current Hubble parameter data was also preliminarily discussed in~\cite{sl5}. In this paper, we wish to perform an uniform analysis for several popular, typical dark energy models, by combining the current $H(z)$ data with the future high-redshift $H(z)$ data from the redshift-drift observations.

The simplest candidate for dark energy is the ``cosmological constant'' $\Lambda$ proposed by Einstein, of which the corresponding cosmological model is the $\Lambda$ cold dark matter ($\Lambda$CDM) model. The $\Lambda$CDM model is very simple and is favored by the current cosmological observations, in particular, the observation of the Planck satellite mission~\cite{Ade:2015xua}, thus it is widely viewed as a prototype of the standard cosmological model. However, actually, current observations have not excluded the dynamical dark energy models, and in fact the $\Lambda$CDM model needs to be tested further in a more accurate manner. Thus it is extremely important to probe the dynamics of dark energy. The simplest extension to $\Lambda$ is the dark energy with a constant $w$, of which the corresponding cosmological model is the so-called $w$CDM model. The shortcoming of this model is that the constant $w$ is usually viewed unphysical or unreal. To consider a model with time-varying $w$, the most popular way is to parametrize $w(a)$ in the form of $w(a)=w_0+w_a(1-a)$, which is often called the Chevallier-Polarski-Linder (CPL) model~\cite{CPL1,CPL2}. However, the CPL model has an evident shortcoming that it has two more additional parameters than $\Lambda$CDM, which adds enormous complexities leading to the fact that $w_0$ and $w_a$ (in particular $w_a$) are very difficult to be well constrained. To remain the same number of parameters with $w$CDM and to simultaneously consider the evolution of $w$, we take the holographic dark energy (HDE) model~\cite{Li:2004rb,Huang:2004mx,Zhang:2014ija} into account. The HDE model originates from the consideration of the holographic principle of quantum gravity, and it can fit the observational data fairly well \cite{Huang:2004wt,Zhang:2005hs,Chang:2005ph,Zhang:2007sh,Li:2009bn,Li:2009jx,Li:2013dha,Wang:2013zca,Zhang:2015rha,Cui:2015oda,Zhang:2015uhk}, thus it is a rather competitive model among the many dark energy models~\cite{Li:2009bn,Li:2009jx,Zhang:2015uhk}. Therefore, in this paper, in order to make a comprehensive analysis, we take the $\Lambda$CDM, $w$CDM, CPL, and HDE models as typical examples.

The organization of this paper is as follows. In Sect. \ref{method}, we describe the current measurements of the Hubble parameter $H(z)$, and introduce the redshift-drift observations from which the high-redshift $H(z)$ data can be obtained. In Sect. \ref{test}, we use the current $H(z)$ data and the simulated $H(z)$ data from the SL test to constrain the typical dark energy models and discuss what role the redshift-drift observations would play in constraining dark energy with the Hubble parameter measurements. Conclusion is given in Sect. \ref{conclusion}.

\section{Method and data}\label{method}

\subsection{The current Hubble parameter measurements}

The Hubble parameter $H(z)$ is defined to be the rate of the relative expansion of the universe,
\begin{equation}\label{2}
    H(z)=\frac{\dot{a}}{a}=-\frac{1}{1+z} \frac{dz}{dt},
\end{equation}
where ${a}$ is the cosmic scale factor and $\dot{a}$ is its rate of change with respect to the cosmic time $t$. $H(z)$ is usually expressed in the unit of km~s$^{-1}$~Mpc$^{-1}$. Directly measuring $H(z)$ is always a major challenge in modern cosmology.

In recent years, enormous efforts have been made in the measurements of $H(z)$. Currently, more than 30 $H(z)$ data have been accumulated, from two kinds of different measurement methods. The first method was proposed by Jimenez and Loeb~\cite{H1} in 2002. One could take the passively evolving galaxies as standard cosmic chronometers whose differential age evolution as a function of the redshift can directly probe $H(z)$, as is given by the second equal sign of Eq. (\ref{2}). This method is usually called {\it differential age} method, abbreviated as ``DA'' method in this paper. We use 25 data obtained from the DA method through more than 10 years' effort, as listed in Table \ref{table1}. These data include eight new measurements of $H(z)$ in $2012$~\cite{H8} with smaller error bars compared to the earlier data~\cite{Chen:2011ys}. In the current literature~\cite{Chen:2011ys,Farooq:2012ju,H10}, it has been shown that the constraints from them on cosmological models are almost equal to those from current type Ia supernova apparent magnitude versus redshift data. Besides, we add two latest $H(z)$ data obtained in 2015~\cite{Moresco:2015cya}, up to $z \sim 2$ ($z = 1.363$ and $z = 1.965$). It has been shown~\cite{Moresco:2015cya} that there is a detectable improvement ($\sim 5\%$) on $\Omega_{m}$ and $w$ compared to previous measurements when they are used to estimate the accuracy on cosmological parameters in the $\Lambda$CDM and $w$CDM models.

\begin{figure*}
\includegraphics[width=7.8cm]{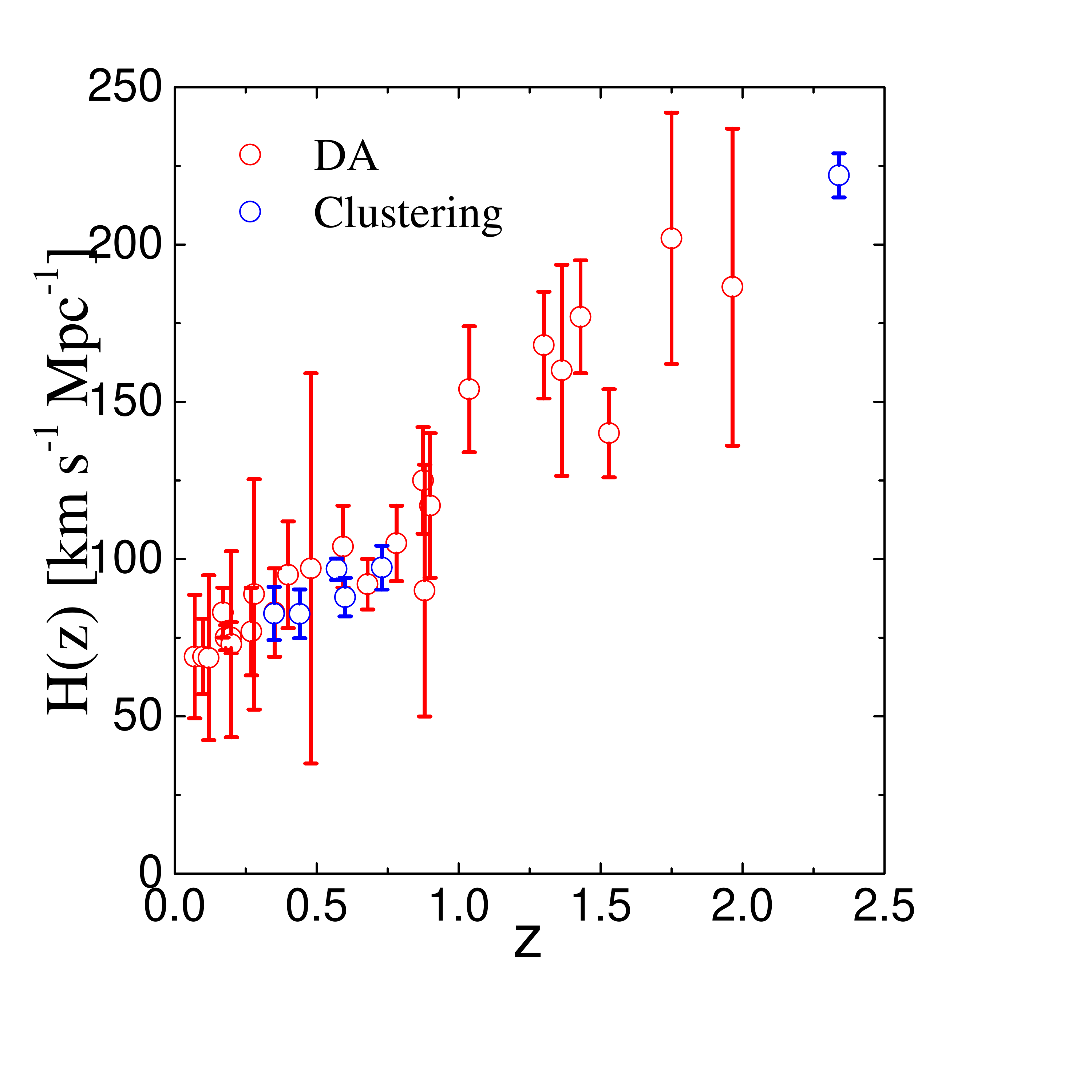}
\includegraphics[width=7.8cm]{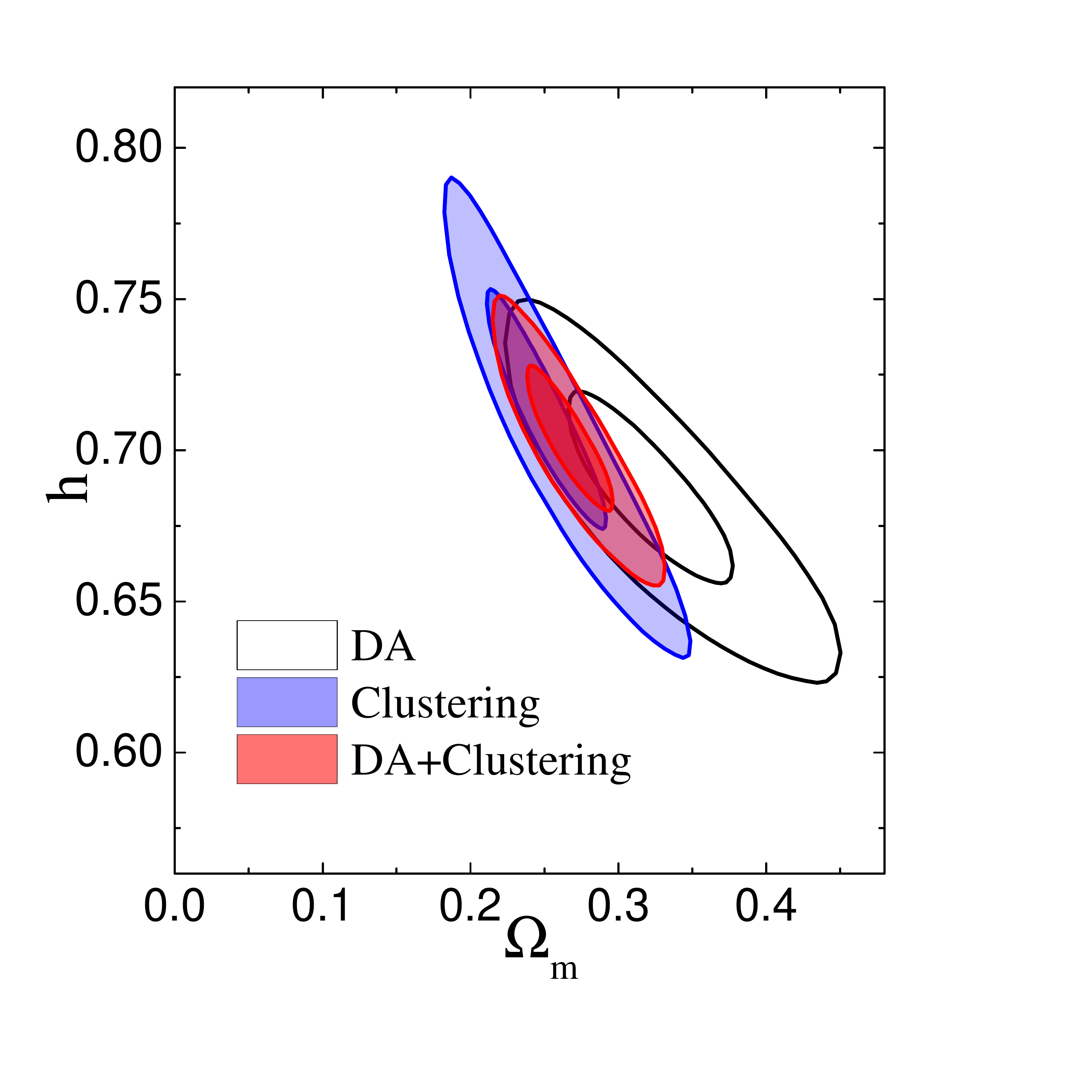}
\caption{\label{fig1} {\it Left}: The current Hubble parameter measurements data (31 points referenced in total), where 25 $H(z)$ data points ($0.07\leq z\leq 1.965$) come from the ``DA'' measurement and six data points ($0.35\leq z\leq 2.34$) come from the ``Clustering'' measurement. In these data points, the highest redshift is $z=2.34$, corresponding to the point obtained from the BAO measurement in the Ly-$\alpha$ forest of BOSS quasars~\cite{Delubac:2014aqe}, in the ``Clustering'' dataset; whereas all the other five points in the ``Clustering'' dataset are in the range of $z\in [0.35, 0.73]$. {\it Right}: The constraints on the $\Lambda$CDM model with the current $H(z)$ measurements (the 68\% and 95\% CL contours are shown in the $\Omega_m$--$h$ plane). We show the constraints from the ``DA'' and ``Clustering'' datasets, separately, and we also show the constraints from the combination of the two.}
\end{figure*}

\begin{table*}\tiny
\caption{Data of the Hubble parameter $H(z)$ versus the redshift $z$, where $H(z)$ and $\sigma_{H}$ are in units of km~s$^{-1}$~Mpc$^{-1}$. %Note that this table is duplicated from Ref. \cite{H15} for a convenient and self-contained discussion in this paper.
}
\label{table1}
\small
\setlength\tabcolsep{2.8pt}
\renewcommand{\arraystretch}{0.8}
\centering
\begin{tabular}{cccccccccc}
\\
\hline\hline
%           \cline{2-4}\cline{6-8}
$z$  & $H(z)$ & $\sigma_{H}$ & Reference & Method  \\ \hline

$0.07$   & $69.0$   & $19.6$    & \cite{H13} & DA \\
$0.1$    & $69.0$    & $12.0$     & \cite{H5} & DA \\
$0.12$   & $68.6$   & $26.2$    & \cite{H13} & DA \\
$0.17$   & $83.0$   & $8.0$    & \cite{H5} & DA \\
$0.179$   & $75.0$   & $4.0$    & \cite{H8} & DA \\
$0.199$   & $75.0$   & $5.0$    & \cite{H8} & DA \\
$0.2$   & $72.9$   & $29.6$    & \cite{H13} & DA \\
$0.27$   & $77.0$   & $14.0$    & \cite{H5} & DA \\
$0.28$   & $88.8$   & $36.6$    & \cite{H13} & DA \\
$0.352$   & $83.0$   & $14.0$    &\cite{H8} & DA \\
$0.4$   & $95.0$   & $17.0$    & \cite{H5} & DA \\
$0.48$   & $97.0$   & $62.0$    & \cite{H5} & DA \\
$0.593$   & $104.0$   & $13.0$    &\cite{H8} &  DA \\
$0.68$   & $92.0$   & $8.0$    & \cite{H8} & DA \\
$0.781$   & $105.0$   & $12.0$    & \cite{H8} & DA \\
$0.875$   & $125.0$   & $17.0$    & \cite{H8} & DA \\
$0.88$   & $90.0$   & $40.0$    & \cite{H5} & DA \\
$0.9$   & $117.0$   & $23.0$    & \cite{H5} & DA \\
$1.037$   & $154.0$   & $20.0$    & \cite{H8} & DA \\
$1.3$   & $168.0$   & $17.0$    & \cite{H5} & DA \\
$1.363$   & $160.0$   & $33.6$    & \cite{Moresco:2015cya} & DA \\
$1.43$   & $177.0$   & $18.0$    & \cite{H5} & DA \\
$1.53$   & $140.0$   & $14.0$    & \cite{H5} & DA \\
$1.75$   & $202.0$   & $40.0$    & \cite{H5} & DA \\
$1.965$   & $186.5$   & $50.4$    & \cite{Moresco:2015cya} & DA \\
$0.35$   & $82.7$   & $8.4$    & \cite{H10} & Clustering \\
$0.44$   & $82.6$   & $7.8$    & \cite{H9} & Clustering \\
$0.57$   & $96.8$   & $3.4$    & \cite{Anderson:2013zyy} & Clustering \\
$0.60$   & $87.9$   & $6.1$    & \cite{H9} & Clustering \\
$0.73$   & $97.3$   & $7.0$    & \cite{H9} & Clustering \\
$2.34$   & $222.0$   & $7.0$    & \cite{Delubac:2014aqe} & Clustering \\
\hline\hline
\end{tabular}
\end{table*}

The second popular way to directly measure $H(z)$ is through the \emph{clustering of galaxies or quasars}. Hereafter, this approach is called ``Clustering" for convenience. One could get a direct measurement of $H(z)$ by using the BAO peak position as a standard ruler in the radial direction \cite{H4}. Through the BAO detection, a measurement of $D_{V} = D^{2/3}_{A} (z/H(z))^{1/3}$ was obtained with a combination of the Hubble parameter $H(z)$ and the angular diameter distance $D_{A}(z)$. Then one can measure $H(z)$ and $D_{A}(z)$ through a variety of scientific methods \cite{H4,H9,H10,Anderson:2013zyy,Delubac:2014aqe}. For example, Gaztanaga et al.~\cite{H4} separated the clustering of the LRG sample in the SDSS DR6 and DR7 into the line-of-sight and transverse information, and obtained $H(z) = 79.69 \pm 2.65 $ km~s$^{-1}$~Mpc$^{-1}$ at $z=0.24$ and $H(z) = 86.45 \pm 3.68 $ km~s$^{-1}$~Mpc$^{-1}$ at $z=0.43$. But the two data are not used in our analysis (consistent with~\cite{Farooq:2012ju,Farooq:2013hq,Farooq:2013eea,Ding:2015vpa}) because they have unreasonably small error bars, causing a strong controversy in the existing papers~\cite{MiraldaEscude:2009uz,Kazin:2010nd,Cabre:2010bc}; Blake et al.~\cite{H9} extracted $H(z)$ and $D_{A}(z)$ by combining the acoustic parameter $A(z) \propto [D^{2}_{A}(z)/H(z)]^{1/3}$ and the Alcock-Paczynski distortion parameter $F(z) \propto D_{A}(z)H(z)$, and obtained $H(z) = 82.6 \pm 7.8 $ km~s$^{-1}$~Mpc$^{-1}$ at $z=0.44$, $H(z) = 87.9 \pm 6.1 $ km~s$^{-1}$~Mpc$^{-1}$ at $z=0.6$, and $H(z) = 97.3 \pm 7.0 $ km~s$^{-1}$~Mpc$^{-1}$ at $z=0.73$; and so on. Detailed separation methods will not be discussed in detail in this paper; for more details, see Refs. \cite{H4,H9,H10,Anderson:2013zyy,Delubac:2014aqe}. Importantly, we use the latest BAO measurement $H(z) = 96.8 \pm 3.4$ km~s$^{-1}$~Mpc$^{-1}$ at $z = 0.57$~\cite{Anderson:2013zyy} and $H(z) = 222 \pm 7$  km~s$^{-1}$~Mpc$^{-1}$ at $z = 2.34$~\cite{Delubac:2014aqe} instead of the previous measurements at the same redshifts. The total six ``Clustering" measurements of $H(z)$ are also listed in Table \ref{table1}.

Note here that in this paper we adopt most of the compilation of the current $H(z)$ data from Ref. \cite{H15}. In Ref. \cite{H15}, the sources of these $H(z)$ data are clearly given, and the statistical and systematical errors are discussed in detail. The other updated $H(z)$ data are also discussed in Refs. \cite{Anderson:2013zyy,Delubac:2014aqe,Moresco:2015cya} in detail. For the utilization of the data from BAO, some authors thought that they are not totally model-independent and thus may not be used in the cosmological parameter constraints~\cite{Daly:2005iu,Liao:2012zza,Liao:2012gq,Melia:2013sxa,Chen:2014tdy,Li:2015nta,Melia:2015nwa,Cai:2015pia}. We admit that there are indeed some problems in the utilization of the $H(z)$ data, but these are not the focus of this paper. The main aim of this paper is to have a look at how the future redshift-drift measurements can improve the constraints on cosmological parameters with the $H(z)$ data alone. In order not to deviate from the main aim of this paper, we do not address these issues in this paper.

We plot these $H(z)$ data points in the left panel of Fig. \ref{fig1}. The 25 data points from the ``DA'' measurement are in the range of $0.07\leq z\leq 1.965$ and the six data points from the ``Clustering'' measurement are in the range of $0.35\leq z\leq 2.34$. For these data points, the highest redshift is $z=2.34$~\cite{Delubac:2014aqe}, corresponding to the point obtained from the BAO measurement in the Ly-$\alpha$ forest of BOSS quasars, in the ``Clustering'' dataset. (Note that using this high-redshift measurement, the evidence of evolving dark energy has been demonstrated in Ref. \cite{Sahni:2014ooa}.) The other five points in the ``Clustering'' dataset are all in the range of $z\in [0.35, 0.73]$. Comparing these data points in Fig.~\ref{fig1}, we apparently find that the error bars of points from the ``Clustering'' dataset are much less than those from the ``DA'' dataset.

In order to constrain the cosmological models with these $H(z)$ data points, we need to perform a $\chi^2$ statistical analysis. The $\chi^2$ function of this analysis is given by
\begin{equation}\label{3}
    \chi^{2}_{H}(p)=\sum^{N}_{i=1}\frac{[H^{\rm th}(z_{i};p)-H^{\rm obs}(z_{i})]^{2}}{\sigma^{2}_{H,i}}~,
\end{equation}
where $N$ denotes the number of data points, $z_{i}$ is the redshift at which $H(z_{i})$ has been measured, $p$ represents model parameters, $H^{\rm th}$ and  $H^{\rm obs}$ are the predicted value of $H(z)$ in the cosmological model and the measured value, respectively, and $\sigma_{H,i}$ is the standard deviation of the $i$th point.

Since the $\Lambda$CDM model is widely viewed as a prototype of the standard cosmology, we take this model as a reference model to test the consistency of the two datasets of $H(z)$ measurements. In the right panel of Fig. \ref{fig1}, we plot the two-dimensional posterior contours (68\% and 95\% confidence level) in the $\Omega_m$--$h$ plane of the $\Lambda$CDM model using the DA and Clustering data of $H(z)$. We find that the two datasets are rather consistent with each other. Although the number of data points is less, the constraining power of the ``Clustering'' set is evidently better than the ``DA'' dataset. The combination of the two datasets provides a much tighter constraint on the cosmological model; see the red contours in this figure. The combined $H(z)$ measurements with 31 data in total give the fit results: $\Omega_m=0.2654^{+0.0325}_{-0.0287}$ and $h=0.7043^{+0.0241}_{-0.0246}$, constraining the parameter $\Omega_m$ to the precision of $\sim 11.57\%$ and the parameter $h$ to the precision of $\sim 3.46\%$. This shows that solely using the current $H(z)$ measurements could provide rather tight constraints on the cosmological parameters.

\subsection{Future high-redshift $H(z)$ measurements from redshift-drift observations}

In this paper, we study how accurate high-redshift $H(z)$ data could be provided by the future redshift-drift observation and how these data would impact on constraining dark energy with the $H(z)$ measurements alone.

The redshift-drift observation, sometimes called the ``SL test'', is not only conceptually simple, but also is a direct probe of cosmic dynamic expansion, although being observationally challenging. We adopt an experiment like CODEX~\cite{Liske:2008ph} to perform a forecast analysis for the predicted accuracy of observations. The major observation facilities, e.g. ELT, aim at directly measuring the accelerating expansion of the universe by detecting the cosmological redshift drift of the Lyman-$\alpha$ forest from QSOs lying in $2 \lesssim z \lesssim 5$.

The main observation of SL test is the redshift variation, expressed as a spectroscopic velocity shift \cite{Loeb:1998bu},
\begin{equation}\label{3}
   \Delta v=\frac{\Delta z}{1+z} =H_{0} \Delta t_{o} \left[1-\frac{E(z)}{1+z}\right],
\end{equation}
where $\Delta t_{o}$ is the time interval of observation, and $E(z)=H(z)/H_{0}$ is decided by specific cosmological models.
According to the performance of the Monte Carlo simulations of Lyman-$\alpha$ absorption lines, the uncertainty on $\Delta v$ can be written as~\cite{Liske:2008ph}
\begin{equation}\label{4}
   \sigma_{\Delta v} = 1.35\left(\frac{S/N}{2370}\right)^{-1}\left(\frac{N_{\rm QSO}}{30}\right)^{-1/2}\left(\frac{1+z_{\rm QSO}}{5}\right)^{x} ~\mathrm{cm}~\mathrm{s}^{-1},
\end{equation}
where $S/N = 3000$ is defined as the spectral signal-to-noise per $0.00125$ nm pixel, $N_{\rm QSO}$ and $z_{\rm QSO}$ are the number and redshift of QSOs, respectively. In addition, the last exponent $x=-1.7$ for $2 \leqslant z \leqslant 4$ and $x = -0.9$ for $z > 4$. In our simulation, 30 SL test data are chosen to uniformly distribute over six redshift bins of $z_{\rm QSO} \in [2,5]$ (namely, the redshift interval $\Delta z=0.5$ for each bin), by observing 30 bright QSOs at high redshifts.

\begin{figure*}
\includegraphics[width=5.0cm]{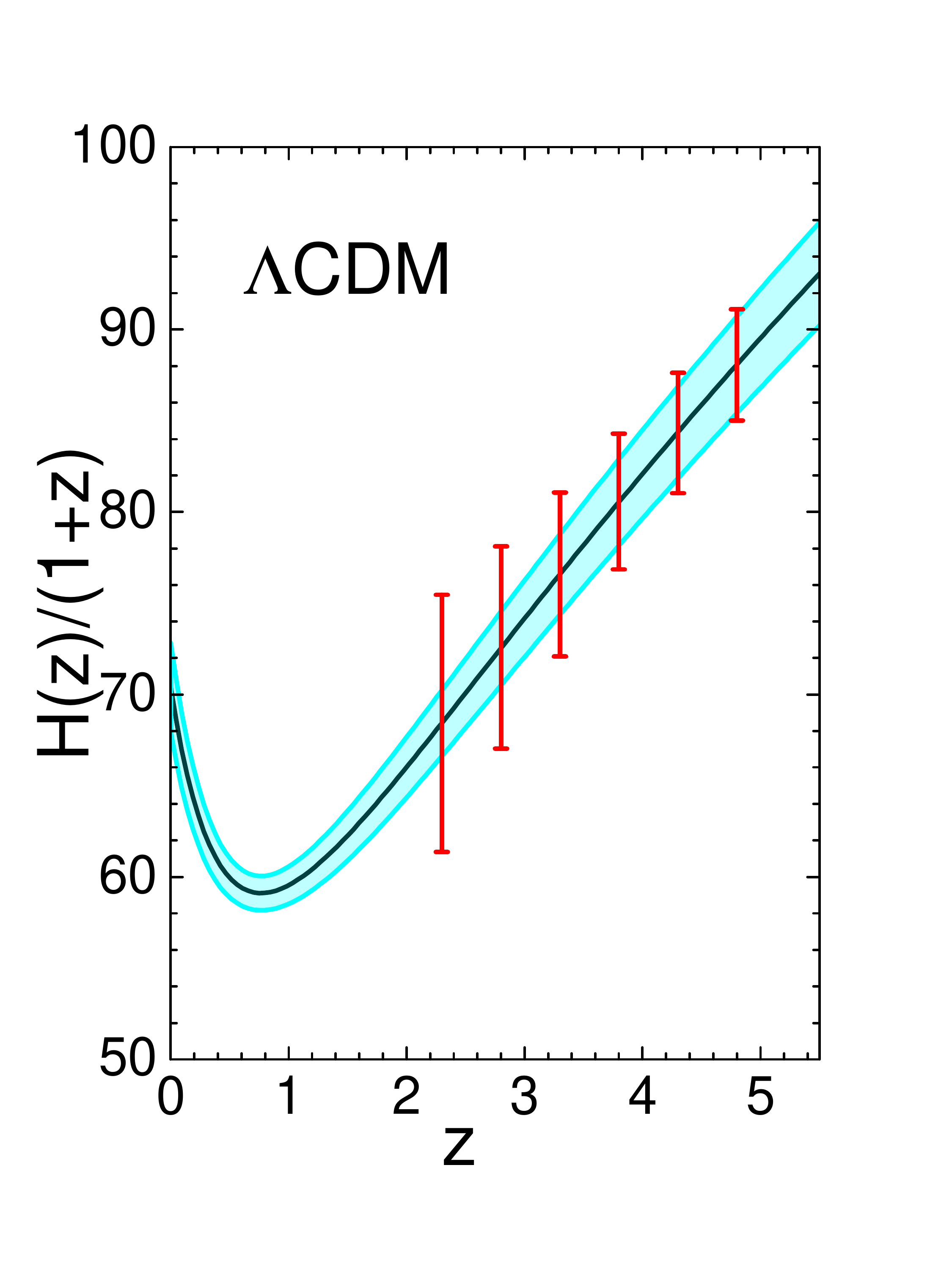}
\includegraphics[width=5.0cm]{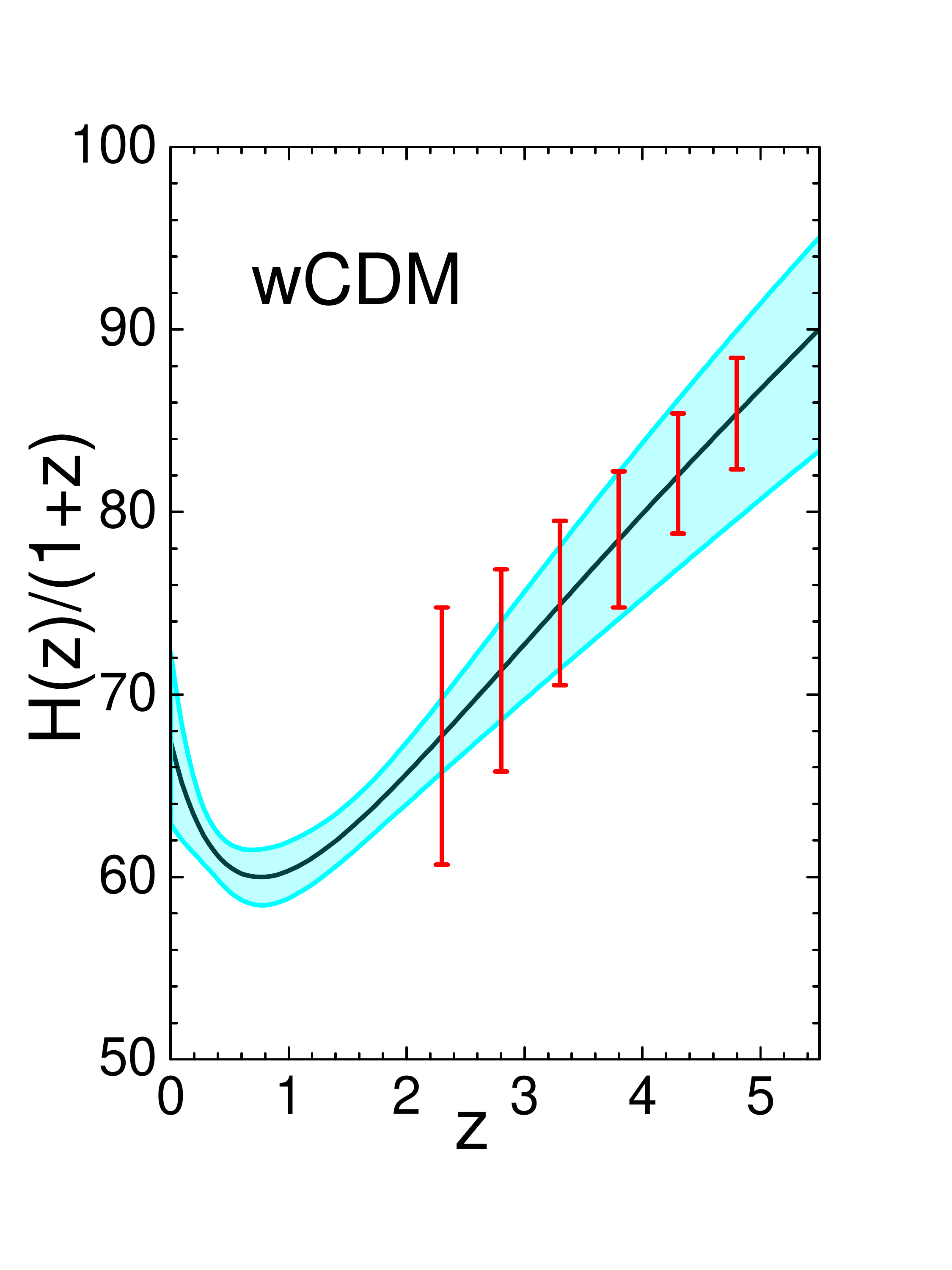}
\includegraphics[width=5.0cm]{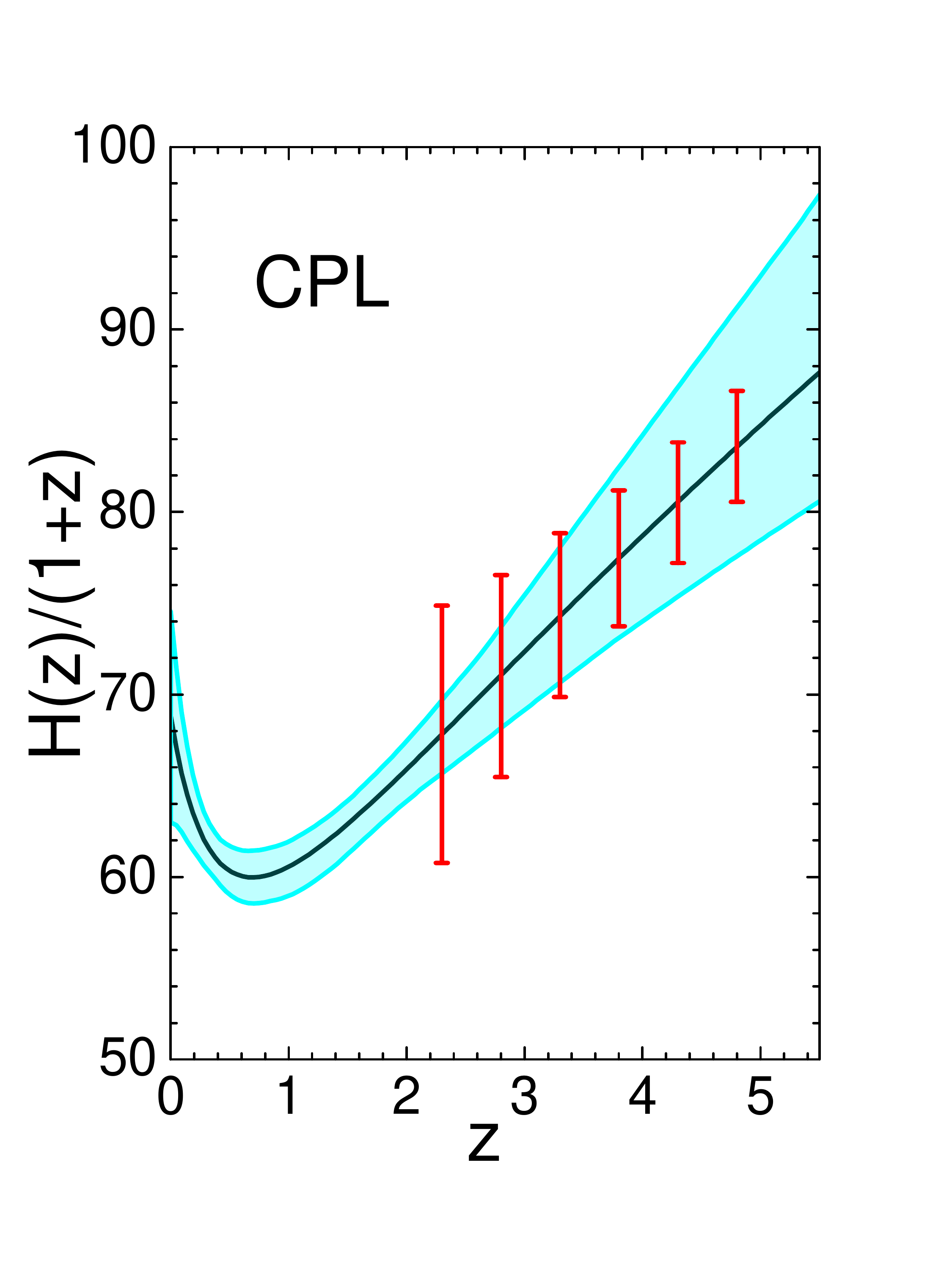}
\caption{\label{fig2} The evolutions of $H(z)/(1+z)$ in the $\Lambda$CDM, $w$CDM, and CPL models. The black curves and cyan bands (best fit with 1$\sigma$ uncertainty) are reconstructed from the current $H(z)$ measurements. The red error bars (1$\sigma$) on the black curves are estimated from the 10-year redshift-drift observation.}
\end{figure*}

The observation of the redshift drift is equivalent to the observation of the Hubble parameter, since we have the simple relationships: $H(z)=(H_0-\Delta v/\Delta t_o)(1+z)$ and $\sigma_H=(1+z)\sigma_{\Delta v}/\Delta t_o$. Thus we can use the SL test to simulate 30 mock $H(z)$ data in the redshift range of $z\in [2, 5]$. In this paper, we choose to consider a 10-year observation of redshift drift ($\Delta t_o=10$ year) to make the analysis, because in our opinion a 10-year forecast is fairly proper and meaningful for our study of the $H(z)$ constraints on dark energy. To show the accuracy of the 10-year $H(z)$ data from the SL test, we plot the forecast data points in Fig. \ref{fig2}; for a convenient display, we show the $H(z)/(1+z)$ plots. The fiducial models for simulating the forecast data are chosen to be the $\Lambda$CDM, $w$CDM, and CPL models in this example, as shown in the three panels of Fig. \ref{fig2}. The black curve with cyan band represents the best fit with 1$\sigma$ uncertainty, reconstructed from the current $H(z)$ measurements (31 data in total). We find that the higher redshift is, the smaller error bar the 10-year SL $H(z)$ point has. This implies that the SL test would play a more important role for constraining the models with more parameters.

In the mock data simulation, we adopt the scheme accordant with our previous papers~\cite{sl4,sl1,sl3,sl2}, i.e., we choose the best-fitting specific dark energy model in study as the fiducial model to produce the simulated $H(z)$ data. The best fit of the dark energy model is given by the current $H(z)$ data. This aims to avoid the potential tension between the current $H(z)$ data and the simulated future $H(z)$ data. In most papers on the redshift-drift observation~\cite{Corasaniti:2007bg,Zhang:2007zga,Balbi:2007fx,Liske:2008ph,Zhang:2010im,Martinelli:2012vq,Li:2013oba,Zhang:2013zyn,sl5}, the fiducial model for simulating data is chosen to be the $\Lambda$CDM model no matter what dark energy model is in study, which sometimes leads to the evident tension between the current data and the simulated data in the combined analysis. Our scheme can efficiently avoid such a problem. In the following, we use the current and future $H(z)$ data to uniformly constrain the typical dark energy models and study what role the high-redshift $H(z)$ measurement from the 10-year SL test would play in constraining dark energy with the $H(z)$ data alone.

\section{Constraints on dark energy models from Hubble parameter measurements including redshift-drift observations}\label{test}

\begin{table*}\tiny
\caption{Fit results for the $\Lambda$CDM, $w$CDM, CPL, and HDE models using the current $H(z)$ data and the $H(z)$ + SL 10-year data.}
\label{table2}
\small
\setlength\tabcolsep{2.8pt}
\renewcommand{\arraystretch}{1.5}
\centering
\begin{tabular}{cccccccccccc}
\\
\hline\hline &\multicolumn{2}{c}{$H(z)$} &&\multicolumn{2}{c}{$H(z)$+ SL 10-year} \\
           \cline{2-3}\cline{5-6}
Parameter  & $\Lambda$CDM & $w$CDM && $\Lambda$CDM & $w$CDM  \\ \hline

$w$                & $-1~({\rm fixed})$
                   & $-0.8174^{+0.2519}_{-0.2563}$&
                   & $-1~({\rm fixed})$
                   & $-0.8151^{+0.1884}_{-0.2391}$\\

$\Omega_{m}$       & $0.2654^{+0.0325}_{-0.0287}$
                   & $0.2662^{+0.0355}_{-0.0376}$&
                   & $0.2654^{+0.0056}_{-0.0055}$
                   & $0.2663^{+0.0093}_{-0.0071}$\\

$\emph{h}$         & $0.7043^{+0.0241}_{-0.0246}$
                   & $0.6742^{+0.0489}_{-0.0451}$&
                   & $0.7042^{+0.0126}_{-0.0126}$
                   & $0.6735^{+0.0423}_{-0.0420}$
                   \\
\cline{2-3}\cline{5-6}
Parameter  & CPL & HDE & & CPL & HDE \\ \hline
$w_{0}$           & $-0.8591^{+0.3981}_{-0.3697}$
                  & $-$&
                  & $-0.8531^{+0.2689}_{-0.2652}$
                  & $-$
                   \\
$w_{a}$            & $0.8583^{+0.4995}_{-2.4411}$
                   & $-$&
                   & $0.8366^{+0.4849}_{-0.4996}$
                   & $-$
                   \\
$c$                & $-$
                   & $1.1383^{+1.3681}_{-0.4323}$&
                   & $-$
                   & $1.1642^{+0.5430}_{-0.3271}$
                   \\

$\Omega_{m}$          & $0.1534^{+0.1799}_{-0.1221}$
                      & $0.2435^{+0.0334}_{-0.0341}$&
                      & $0.1535^{+0.0094}_{-0.0071}$
                      & $0.2421^{+0.0102}_{-0.0146}$
                      \\

$\emph{h}$              & $0.6884^{+0.0571}_{-0.0585}$
                        & $0.6796^{+0.0536}_{-0.0485}$&
                        & $0.6881^{+0.0493}_{-0.0499}$
                        & $0.6785^{+0.0257}_{-0.0247}$
                        \\
\hline\hline
\end{tabular}
\end{table*}

\begin{table*}\tiny
\caption{Constraint errors and precisions of parameters in the $\Lambda$CDM, $w$CDM, CPL, and HDE models for the fits to the current
$H(z)$ data and the $H(z)$ + SL 10-year data.}
\label{table3}
\small
\setlength\tabcolsep{2.8pt}
\renewcommand{\arraystretch}{1.5}
\centering
\begin{tabular}{cccccccccccc}
\\
\hline\hline &\multicolumn{4}{c}{$H(z)$} &&\multicolumn{4}{c}{$H(z)$+ SL 10-year} \\
           \cline{2-5}\cline{7-10}
Error  & $\Lambda$CDM & $w$CDM & CPL & HDE&& $\Lambda$CDM & $w$CDM & CPL  & HDE \\ \hline

$\sigma(w_{0})$     & $-$
                    & $0.2541$
                    & $0.3842$
                    & $-$&
                    & $-$
                    & $0.2152$
                    & $0.2671$
                    & $-$
                    \\

$\sigma(w_{a})$    & $-$
                   & $-$
                   & $1.7619$
                   & $-$&
                   & $-$
                   & $-$
                   & $0.4923$
                   & $-$
                   \\

$\sigma(c)$        & $-$
                   & $-$
                   & $-$
                   & $1.0145$&
                   & $-$
                   & $-$
                   & $-$
                   & $0.4482$       \\

$\sigma(\Omega_{m})$    & $0.0307$
                        & $0.0366$
                        & $0.1537$
                        & $0.0338$&
                        & $0.0056$
                        & $0.0083$
                        & $0.0083$
                        & $0.0126$  \\

$\sigma(\emph{h})$      & $0.0244$
                        & $0.0470$
                        & $0.0578$
                        & $0.0511$&
                        & $0.0126$
                        & $0.0422$
                        & $0.0496$
                        & $0.0252$\\
\cline{2-5}\cline{7-10}
Precision  & $\Lambda$CDM & $w$CDM & CPL & HDE&& $\Lambda$CDM & $w$CDM & CPL  & HDE \\ \hline
$\varepsilon(w_{0})$     & $-$
                    & $0.3109$
                    & $0.4472$
                    & $-$&
                    & $-$
                    & $0.2640$
                    & $0.3130$
                    & $-$
                    \\

$\varepsilon(w_{a})$    & $-$
                   & $-$
                   & $2.0528$
                   & $-$&
                   & $-$
                   & $-$
                   & $0.5885$
                   & $-$
                   \\

$\varepsilon(c)$        & $-$
                   & $-$
                   & $-$
                   & $0.8912$&
                   & $-$
                   & $-$
                   & $-$
                   & $0.3850$       \\

$\varepsilon(\Omega_{m})$    & $0.1157$
                        & $0.1375$
                        & $1.0020$
                        & $0.1388$&
                        & $0.0211$
                        & $0.0312$
                        & $0.0541$
                        & $0.0520$  \\

$\varepsilon(\emph{h})$      & $0.0346$
                        & $0.0697$
                        & $0.0840$
                        & $0.0752$&
                        & $0.0179$
                        & $0.0627$
                        & $0.0721$
                        & $0.0371$ \\
\hline\hline
\end{tabular}
\end{table*}

\begin{figure*}
\includegraphics[width=16.0cm]{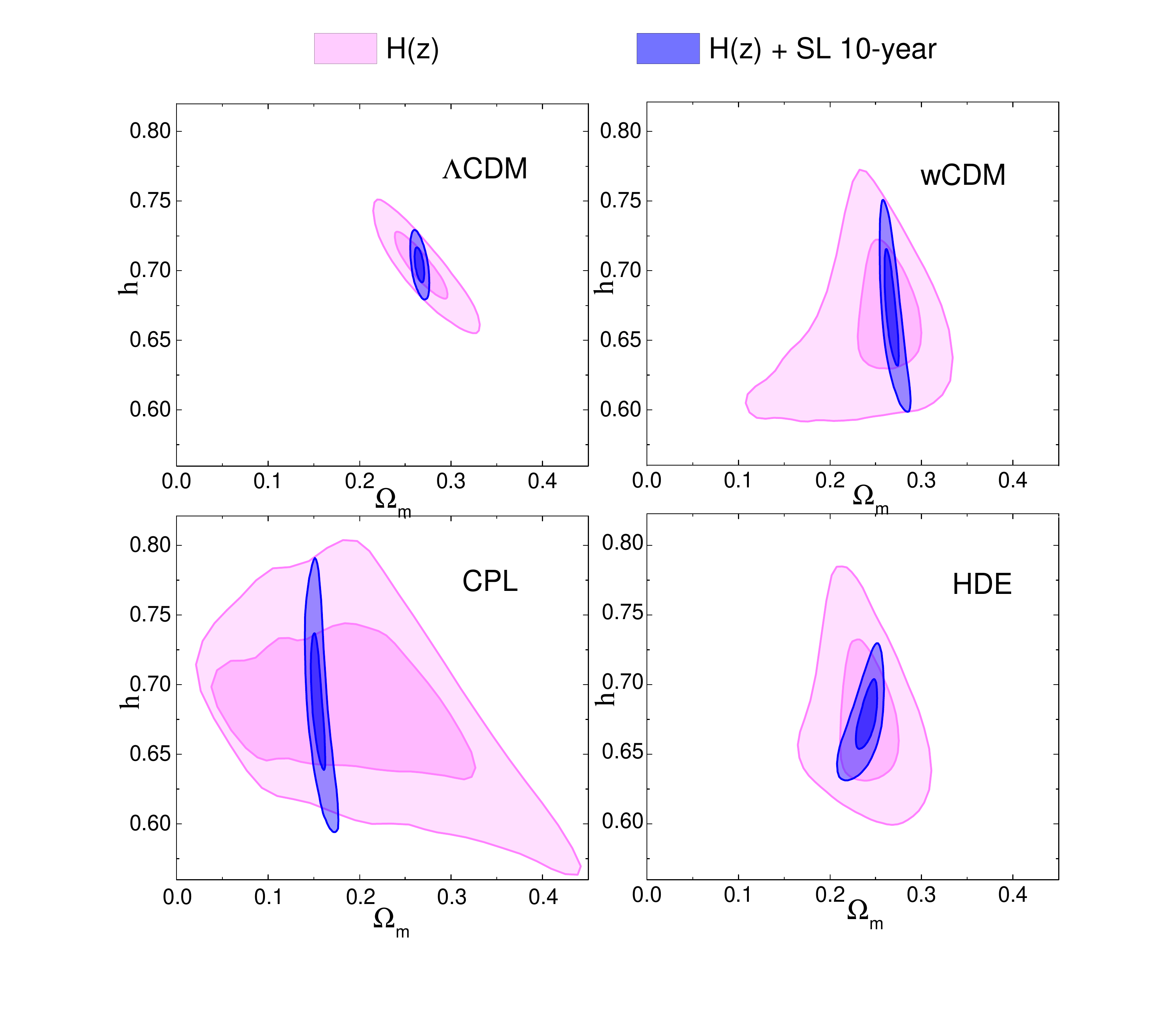}
\caption{\label{fig3} Constraints ($1\sigma$ and $2\sigma$ CL) on the $\Lambda$CDM, $w$CDM, CPL, and HDE models in the $\Omega_m$--$h$ plane from the current $H(z)$ measurements (\emph{pink contours}) and current $H(z)$ + future 10-year redshift-drift measurements (\emph{blue contours}). }
\end{figure*}

\begin{figure*}
\includegraphics[width=16.0cm]{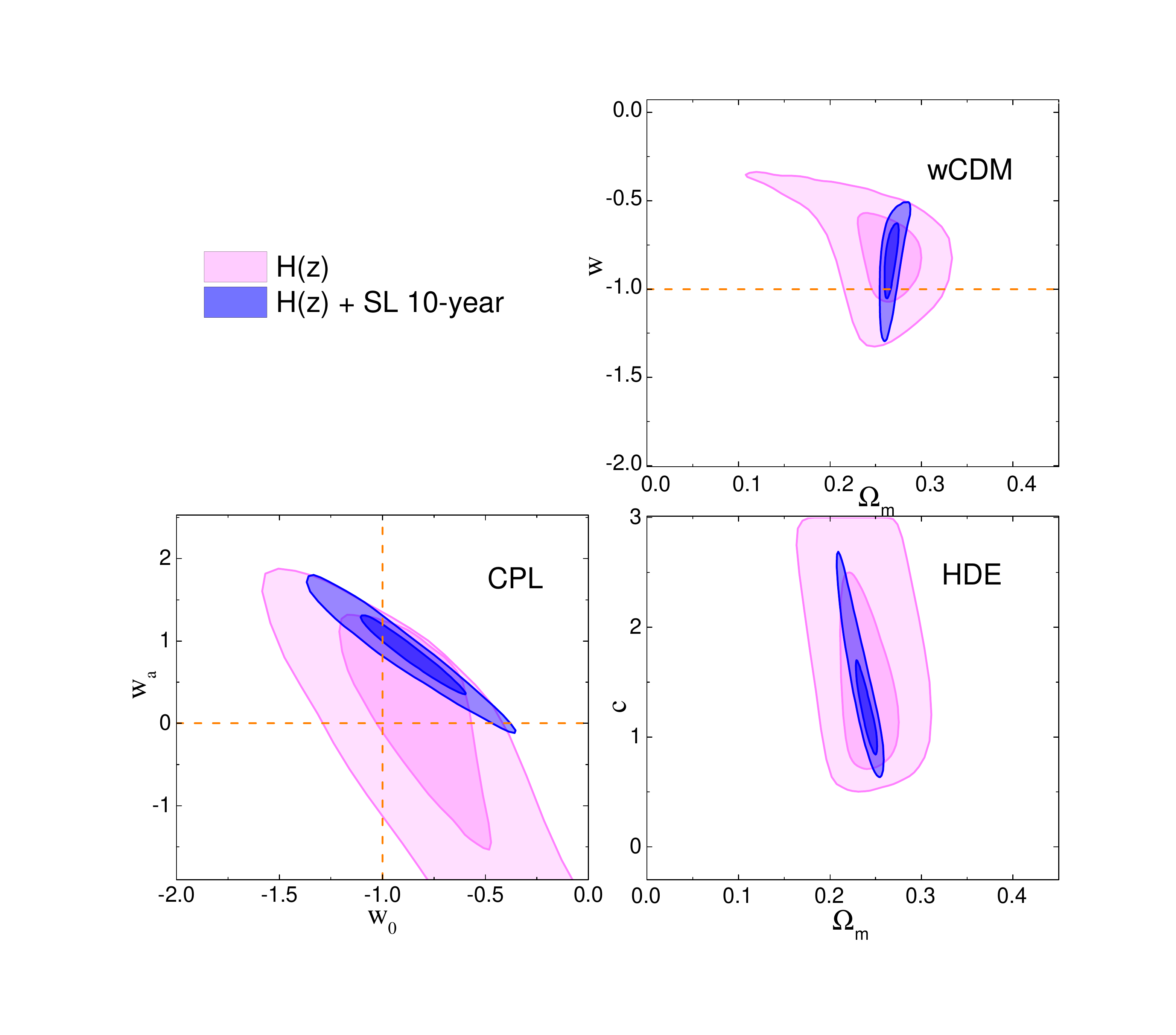}
\caption{\label{fig4} Constraints ($1\sigma$ and $2\sigma$ CL) on the $w$CDM, CPL, and HDE models from the current $H(z)$ measurements (\emph{pink contours}) and current $H(z)$ + future 10-year redshift-drift measurements (\emph{blue contours}). We show the two-dimensional marginalized contours in the $\Omega_m$--$w$ plane for the $w$CDM model, in the $w_0$--$w_a$ plane for the CPL model, and in the $\Omega_m$--$c$ plane for the HDE model. For the models that contain $\Lambda$CDM as a sub-model, namely $w$CDM and CPL, the positions of the cosmological constant are clearly denoted.}
\end{figure*}

\begin{figure*}
\includegraphics[width=7.8cm]{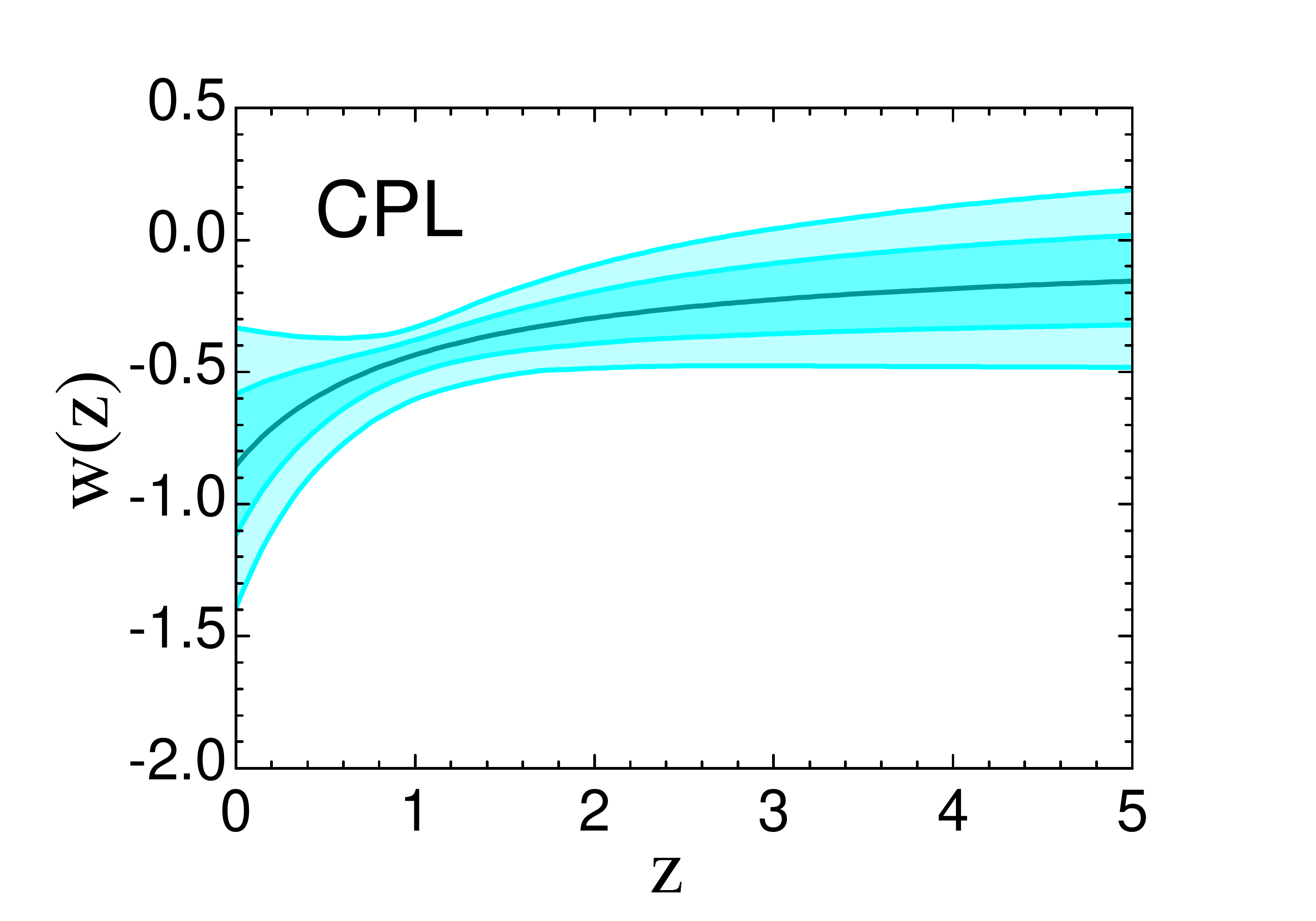}
\includegraphics[width=7.8cm]{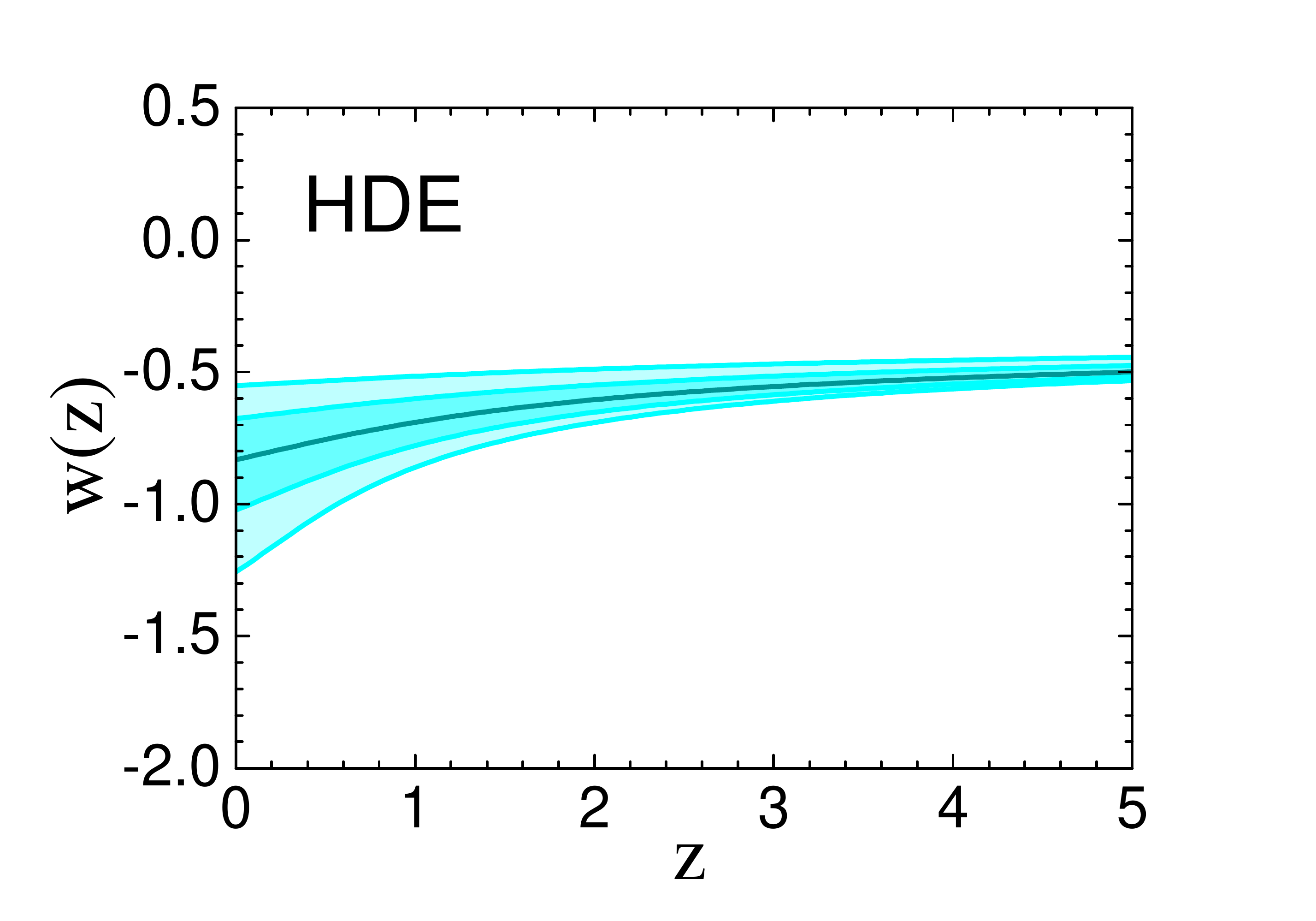}
\caption{\label{fig5} The reconstructed evolutions of $w(z)$ for CPL and HDE with errors ($1\sigma$ and $2\sigma$) obtained from the $H(z)$+ SL 10-year data. }
\end{figure*}

In the section, we study the capability of the $H(z)$ measurements in constraining dark energy models. First, we study how the current 31 $H(z)$ data can be used to constrain the typical dark energy models. Then we use the each best-fitting dark energy model itself as the fiducial model to produce the simulated mock high-redshift $H(z)$ data ($z\in [2, 5]$) from a 10-year redshift-drift observation and combine the current and future $H(z)$ data to constrain the dark energy model. Our aim is to see how the future high-redshift $H(z)$ measurements from SL test would improve the constraining power in the study of dark energy with the $H(z)$ observations.

We choose four specific dark energy models as representatives of cosmological models to make the analysis. They are the $\Lambda$CDM, $w$CDM, CPL, and HDE models. In the $\Lambda$CDM model, the EoS of dark energy is fixed to be $w=-1$. In the $w$CDM model, the EoS of dark energy, $w$, is a constant. In the CPL model, the EoS of dark energy is parametrized as $w(z)=w_0+w_a \frac{z}{1+z}$~\cite{CPL1,CPL2}. In the HDE model, the EoS of dark energy is given by $w(z)=-1/3-(2/3c)\sqrt{\Omega_{\rm de}(z)}$~\cite{Li:2004rb}, where $c$ is a dimensionless parameter and the function $\Omega_{\rm de}(z)$ is the solution to the differential equation $\Omega_{\rm de}'=\Omega_{\rm de}(1-\Omega_{\rm de})[1+(2/c)\sqrt{\Omega_{\rm de}}]$ \cite{Li:2004rb}, with the prime denoting the derivative with respect to $\ln a$.

We constrain the four dark energy models by using the current $H(z)$ data and the combination of the current and SL 10-year $H(z)$ data. The fit results are given in Table \ref{table2}. We find that, using the current $H(z)$ data only, the $\Lambda$CDM model can be well constrained, but other dark energy models that have one or two more parameters than $\Lambda$CDM can only be loosely constrained. However, when the SL 10-year $H(z)$ data are combined, all the constraint results are improved significantly.

To see the improvements from the SL 10-year measurement visually, we show the constraint results in Figs. \ref{fig3} and \ref{fig4}. In Fig. \ref{fig3}, we show the two-dimensional posterior distribution contours (68\% and 95\% CL) in the $\Omega_m$--$h$ plane for the four dark energy models. The pink contours are from the constraints of current $H(z)$ data and the blue contours are from the constraints of the combination of current and SL 10-year $H(z)$ data. For all the cases, we find that the degeneracy directions are evidently changed by adding the SL 10-year data. In Fig. \ref{fig4}, we show the two-dimensional marginalized contours in the $\Omega_m$--$w$ plane for the $w$CDM model, in the $w_0$--$w_a$ plane for the CPL model, and in the $\Omega_m$--$c$ plane for the HDE model. From these figures, we clearly see that adding the SL 10-year data leads to significant improvements for the constraint precisions of all the parameters, in particular the parameter $\Omega_m$.

%In Fig. \ref{fig4}, we show the cosmological constant solution (dashed orange lines) in the contour plots of equation of state. For the $w$CDM model, the cosmological constant $w = -1$ traverses the $1 \sigma$ contour whether by using only $H(z)$ data or $H(z)$+ SL 10-yr data. But for the CPL model, the results from $H(z)$ data are in better agreement with a cosmological constant point ($w_{0}, w_{a}$) = ($-1, 0$) than those from only $H(z)$+ SL 10-yr data combination, whose cosmological constant point lies within the $2 \sigma$ CL.

In order to quantify the improvements, we list the errors and constraint precisions of parameters in the four models for the fits to the current $H(z)$ data and the current + SL 10-year $H(z)$ data, in Table \ref{table3}. %Due to the fact that the fit results are not in the form of totally normal distributions, we define the error as $\sigma=\sqrt{\sigma_+^2+\sigma_-^2\over 2}$, where $\sigma_+$ and $\sigma_-$ are the 1$\sigma$ deviations of upper and lower limits, respectively.
Based on the best-fit value and the error of the parameter in the fit, we can evaluate the constraint precision of the parameter. For a parameter $\xi$, one can define the constraint precision as $\varepsilon(\xi)=\sigma(\xi)/\xi_{\rm bf}$, where $\xi_{\rm bf}$ denotes the best-fit value of $\xi$. %We thus list the constraint precisions of parameters in dark energy models from the two datasets [$H(z)$ and $H(z)$+SL 10-yr], in Table \ref{table3}.

We find that the precision of $\Omega_m$ can be enhanced by nearly one order of magnitude when the SL 10-year $H(z)$ data are combined. Concretely, the precision of $\Omega_m$ is improved from 11.57\% to 2.11\% for $\Lambda$CDM, from 13.75\% to 3.12\% for $w$CDM, from 100.20\% to 5.41\% for CPL, from 13.88\% to 5.20\% for HDE. %For the CPL model, solely using the current $H(z)$ data cannot well constrain $\Omega_m$ (can only give an upper limit), but when the SL data are added, $\Omega_m$ can be constrained to 3.36\%.
The constraint precision of the parameter $h$ is also evidently enhanced for all the four models; for details, see Table \ref{table3}. For the property of dark energy, the constraints are also improved moderately. For example, in the $w$CDM model, the precision of $w$ is improved from 31.09\% to 26.40\%; in the CPL model, the precision of $w_0$ is improved from 44.72\% to 31.30\%; and in the HDE model, the precision of $c$ is improved from 89.12\% to 38.50\%. In addition, for the CPL model, $w_a$ is loosely constrained by only using the current $H(z)$ data (its precision is 205.28\%), but when the SL 10-year observation is considered, $w_a$ can be constrained to 58.85\%. Hence, we see that the $H(z)$ constraints on dark energy models can be improved greatly when the $H(z)$ data from an only 10-year observation of redshift drift are included.

%Furthermore, in Fig. \ref{fig5}, we construct the $w(z)$ evolutions in the CPL and HDE models by using the constraint results of the current $H(z)$ + SL 10-yr data. We see that the CPL model with 2 parameters ($w_{0}, w_{a}$) would produce bigger errors at high redshifts, resulting in loose constraints on the cosmological parameters. Instead, for the HDE model with the single parameter ($c$), the $w(z)$ evolution is rather stable , and has small errors at high redshifts. Hence, the CMB data at high redshifts could make great improvements of parameters constraint in the CPL model. Certainly, it is evident that the nature of dark energy from the HDE model can be described by the CPL model parameterized in Fig. \ref{fig5}.

Among the four dark energy models analyzed in this paper, the CPL and HDE models could describe time-evolving EoS $w(z)$. For the CPL model, there are two parameters, $w_0$ and $w_a$, describing the property of dark energy, and the $\Lambda$CDM model is contained in this model as a sub-model with $(w_0,w_a)=(-1,0)$. The HDE model is totally different from the CPL model; it has only one parameter (namely, $c$) to describe the property of dark energy, $w(z)=-1/3-(2/3c)\sqrt{\Omega_{\rm de}(z)}$~\cite{Li:2004rb}. Clearly, in the early times ($z\rightarrow \infty$ and $\Omega_{\rm de}\rightarrow 0$), one has $w(z\rightarrow \infty)=-1/3$, and in the far future ($z\rightarrow -1$ and $\Omega_{\rm de}\rightarrow 1$), one has $w(z\rightarrow -1)=-1/3-2/3c$; thus the HDE model does not involve the $\Lambda$CDM model. In Fig. \ref{fig5}, we show the reconstructed evolutions of $w(z)$ for CPL and HDE with $1\sigma$ and $2\sigma$ errors obtained from the $H(z)$+SL 10-year data. We find that it is possible to differentiate dynamical dark energy from $\Lambda$CDM by only using the $H(z)$ measurements in the future.

\section{Conclusion}\label{conclusion}

The direct measurements of the Hubble parameter at different redshifts are vitally important for constraining the property of dark energy. Usually, the constraints on dark energy are often provided by the distance-redshift relation measurements, but the distance (luminosity distance or angular diameter distance) is linked to dark energy by an integral over $1/H(z)$, and $H(z)$ is affected by dark energy via another integral over $w(z)$. Thus using the distance measurements to constrain the history of $w(z)$ is extremely difficult, but using the $H(z)$ measurements to constrain the dark energy is much simpler and more feasible.

Though directly measuring $H(z)$ is a challenging task, in recent years some $H(z)$ data have been accumulated through the great efforts of astronomers. Up to now, we have about 31 $H(z)$ data in total, covering the redshift range of $z\in [0.07, 2.34]$. In these data, about 25 data points come from the ``DA'' measurement ($0.07\leq z\leq 1.965$) and about six data points come from the ``Clustering'' measurement ($0.35\leq z\leq 2.34$). We show that the two datasets of $H(z)$ are consistent with each other, and solely using the current $H(z)$ data (the combination of the two datasets) could provide fairly good constraints on the typical dark energy models.

In addition, the future redshift-drift observations (i.e., the SL test) could actually also directly measure $H(z)$ at higher redshifts, covering the redshift range of $z\in [2, 5]$. Thus we also discuss what role the redshift-drift observation can play in constraining dark energy with the Hubble parameter measurements. We choose four specific dark energy models as typical examples to make an analysis. They are the $\Lambda$CDM, $w$CDM, CPL, and HDE models. We consider a 10-year observation of redshift drift and produce 30 simulated $H(z)$ data at the redshift range of $z\in [2, 5]$. We show that the constraints on the dark energy models can be improved greatly when the high-redshift $H(z)$ data from only a 10-year observation of redshift drift are combined. We expect that the redshift-drift observation would be successfully implemented and the accurate high-redshift $H(z)$ data could be obtained to make great contribution to the study of dark energy.

\begin{acknowledgments}
We thank Jia-Jia Geng, Yun-He Li, Fulvio Melia, Bharat Ratra, Jing-Fei Zhang, and Ming-Ming Zhao for helpful discussions.
This work is supported by the Top-Notch Young Talents Program of China, the National Natural Science Foundation of China (Grants No.~11522540 and No.~11175042), and the Fundamental Research Funds for the Central Universities (Grants No. N140505002, No. N140506002, and No. L1505007).
\end{acknowledgments}

\end{document}